\documentstyle[aps,prl,epsf,multicol]{revtex}

\begin{document}
\draft

\title{Optical conductivity in doped manganites with planar x$^2$-y$^2$
orbital order} 

\author{Frank Mack and Peter Horsch}

\address{
Max-Planck-Institut f\"{u}r Festk\"{o}rperforschung,
Heisenbergstr.~1, D-70569 Stuttgart (Germany)}

\date{22 December 1998}
\maketitle

\begin{abstract}
We investigate a planar model for the ferromagnetic (FM) phase of manganites,
which develops orbital order of $e_g$ electrons with x$^2$-y$^2$-symmetry at
low temperature. The dynamic structure factor of orbital excitations
and the optical conductivity
$\sigma(\omega)$ are studied with help of a finite-temperature diagonalization
method.
Our calculations provide a theoretical prediction for $\sigma(\omega)$ for the
2D FM state and are 
of possible relevance for the recently found A-type phase of manganites
at high doping which consists of FM layers coupled antiferromagnetically.
In the  x$^2$-y$^2$ ordered regime  $\sigma(\omega)$ shows both a 
Drude peak and a gapped incoherent absorption due to a gap
in the orbital excitations. 
\end{abstract}

\begin{multicols}{2}
Recently Kawano  {\it et al.}\cite{Kawano97} and
Akimoto {\it et al.}\cite{Akimoto98} observed 
a new ground state in heavily doped 3D manganites
Pr$_{1/2}$Sr$_{1/2}$MnO$_3$ and (La$_{1-z}$Nd$_z$)$_{1-x}$Sr$_x$MnO$_3$
(e.g. $x=0.54$ and $z=0.0-1.0$), which surprisingly 
show the same antiferromagnetic A-structure known from the undoped 
LaMnO$_3$
where ferromagnetic (FM) planes are coupled antiferromagnetically. It was 
proposed that the occupied orbitals in this phase have x$^2$-y$^2$ 
symmetry\cite{Akimoto98} in contrast to the alternating orbital order
in LaMnO$_3$\cite{Murakami98}, and
as a consequence the conduction should be quasi-2D and the magnetic coupling
between the layers small.
This broken symmetry state has to be distinguished from the orbital structure 
of the 3D FM state realized at lower doping, where orbital correlations
are expected to be liquid like\cite{Ishihara97b,Kilian98}.  
The latter FM state shows a highly anomalous form of the optical conductivity
\cite{Okimoto95,Kim98,Quijada98} with a large incoherent absorption extending 
up to $\omega\sim 1$eV.
It has been argued that the
anomalous incoherent absorption is related to the orbital degree of freedom
\cite{Ishihara97b,Kilian98,Shiba97} and characteristic for an {\em orbital liquid}
\cite{Ishihara97b,Kilian98}. However there are also alternative scenarios
which attribute the incoherent feature to Jahn-Teller 
polarons\cite{Kim98,Quijada98}.
Comparable optical studies in the high doping regime where the 
{\em orbital ordered} A-phase is
the stable ground state have not yet been reported. The purpose of this
work is to present a theoretical analysis of the possible outcome of such
experiments.  The frequency dependence of $\sigma(\omega)$ is interesting since it is
expected to reveal the orbital excitations in the doped   x$^2$-y$^2$ ordered phase,
and may allow to distinguish between the {\em orbital excitation} and
the {\em Jahn-Teller polaron} scenario. 
\newline\indent
The evolution of x$^2$-y$^2$ orbital order upon doping was also found in the
study of planar models derived from the degenerate Kondo lattice model
in the  fully polarized FM phase\cite{Horsch98}.
Therefore we believe that this model is an appropriate starting point
to investigate these questions.
The $t_{2g}$-spins of the Mn$^{3+}$ and Mn$^{4+}$ ions as well the 
$e_g$-electron spins of  Mn$^{3+}$ are aligned globally in the FM-phase 
and can be integrated out. What remains is the orbital degree of freedom since
there are two nearly degenerate $e_g$ orbitals per site.
The large intra-atomic repulsive interactions prevent double occupancy of
$e_g$ orbitals at the same site, which leads to a strong correlation 
problem and to nontrivial charge and orbital dynamics.
\newline\indent
For the saturated FM state
the restriction to configurations without double occupancy leads to the
{\em orbital} $t$-$J$ model\cite{Ishihara97a,Horsch98}
\begin{equation}
H_{orb}=-\sum_{\langle {\bf ij} \rangle ab} (t_{\bf ij}^{ab}
\tilde{d}_{{\bf i} a}^{\dagger} \tilde{d}_{{\bf j} b}+ H.c.)+H_{int}.
\label{eq:Hfm}
\end{equation}
with 
${\tilde d}^{\dagger}_{{\bf i}a}=d^{\dagger}_{{\bf i}a}(1-n_{{\bf i}{\bar a}})$.
Here we use $a$ and $b$ ($\alpha$ and $\beta$)
as orbital-pseudospin indices; while 
${\bar a}$ denotes the orthogonal $e_g$ orbital with respect to orbital $a$.
A convenient basis is $|\uparrow\rangle=d_{x^2-y^2}$ and 
$|\downarrow\rangle =d_{3z^2-r^2}$. The transfer
matrix elements are then given by
\begin{equation}
t_{{\bf ij}\parallel x/y}^{ab}=\frac{t}{4}\left( \begin{array}{cc}
3 & \mp\sqrt{3}\\
\mp\sqrt{3} & 1\\
\end{array} \right),\;\;
\end{equation}
which allows for inter-orbital hopping in the $xy$-plane. The
$\mp$-sign distinguishes hopping along $x$ and $y$ direction.
Here $t/4$ is the matrix element
between $d_{3z^2-r^2}$ orbitals in the $xy$-plane.
For one electron per site this model describes a Mott insulator. 
The orbital interaction $H_{int}$ follows as a consequence of the 
elimation of doubly occupied sites with energy of an effective $U$\cite{U},
where we assume that because of the large Hund-splitting
the high-spin intermediate states are most important. 
\begin{eqnarray}
H_{int} = -\sum_{\bf j u u'}\sum_{a b \alpha \beta}
&&\frac{t^{\alpha \beta}_{\bf j+u\; j}t^{b a}_{\bf j \; j+u'}}{U}
\Bigl[\delta_{\beta,b}\;  \tilde{d}^{\dagger}_{{\bf j+u} \alpha}
\tilde{d}^{\dagger}_{{\bf j} \bar{b}} \tilde{d}_{{\bf j} \bar{b}}
\tilde{d}_{{\bf j+u'} a}  \nonumber\\
&-&\delta_{\beta \bar{b}} \tilde{d}^{\dagger}_{{\bf j+u} \alpha}
\tilde{d}^{\dagger}_{{\bf j} b} \tilde{d}_{{\bf j}\bar{b} }
\tilde{d}_{{\bf j+u'} a}  \Bigr] .
\end{eqnarray}
Here ${\bf u, u'}=(\pm a,0)$ or $(0,\pm a)$ are lattice unit vectors.
The orbital interaction  $H_{int}=H^{(2)}_{int}+H^{(3)}_{int}$
consists of two types of
contributions: (i) 2-site terms
(${\bf u}={\bf u'}$), i.e. similar to the Heisenberg interaction in the
standard $t$-$J$ model, yet more complex because of the nonvanishing 
off-diagonal $t^{ab}$, and (ii) 3-site hopping terms
(${\bf u} \neq {\bf u'}$) between second nearest neighbors.
In the half-filled case only the
2-site interaction is operative and may induce some orbital order.
In the presence of hole doping, i.e. for less than one $e_g$-electron
per site, both the kinetic energy and the 3-site contributions in
$H_{int}$ lead to propagation of the holes and to a frustration of
the orbital order prefered by the orbital interaction $H^{(2)}_{int}$.
When neglecting the 3-site terms our Hamiltonian is equivalent to the 
orbital model studied by Ishihara {\it et al.}\cite{Ishihara97a}.
Since the spectral shape of $\sigma(\omega)$ is only weakly affected
by these terms\cite{Horsch98}, we do not consider them in this study.
For the calculation of correlation functions (CF) 
we use a generalization of
the exact diagonalization technique for finite temperature developed
by Jakli\v c and Prelov\v sek\cite{Jaklic94}.  
%
\begin{figure}
\epsfxsize=5.6cm
\centerline{\hbox{\epsffile[60 20 530 680]{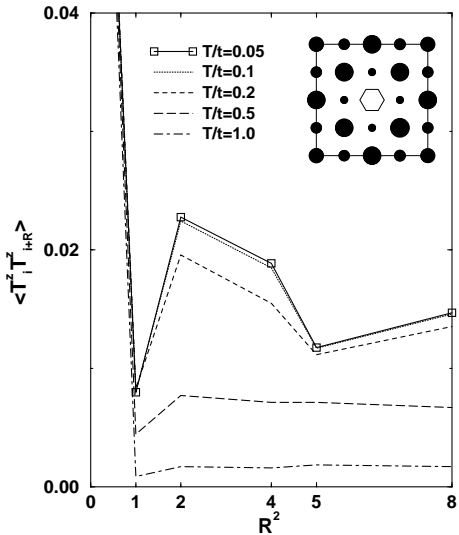}}}
\noindent
\caption{ \label{fig1}
Evolution of orbital correlations $\langle T^z_{\bf i}T^z_{\bf i+R}\rangle$
as function of distance $R$ for various 
temperatures $T$. The ratio of hole number and system size is N$_h$/N=10/16 and $U=16t$.
The inset gives a schematic view of the orbital correlations for $T=0.05 t$
(strength proportional to radius). 
}
\end{figure}

Figure 1 monitors the evolution of $x^2$-$y^2$ orbital correlations as 
function of distance and temperature for a 16-site cluster\cite{BC}. 
Here we use the pseudospin operator notation
$T_{\bf i}^z=\frac{1}{2}(n_{{\bf i}\uparrow}-n_{{\bf i}\downarrow})$,
$T_{\bf i}^+=\tilde{d}^{\dagger}_{{\bf i}\uparrow}\tilde{d}_{{\bf i}\downarrow}$
and
$T_{\bf i}^-=\tilde{d}^{\dagger}_{{\bf i}\downarrow}\tilde{d}_{{\bf i}\uparrow}$.
There is a clear manifestation of 
long-range orbital correlations below $T=0.5 t$ which saturates 
below $T=0.2 t$.
Although $\langle T^z_{\bf i}T^z_{\bf i+R}\rangle\sim 0.02$ appears small,
it is close the maximal value $\sim 0.03$ for this CF
given the electron concentration $6/16$.
This CF has a  minimum for nearest-neighbors
(see inset of Fig.1),   which is 
partially due to the evolution of an exchange-correlation hole in the
charge CF $\langle n_{\bf i}n_{\bf i+R}\rangle$. 
In addition $\langle n_{\bf i}n_{\bf i+R}\rangle$ shows further neighbor density
modulations, which disappear with the orbital order. 
Usually such charge correlations
near quarter-filling are attributed to strong nearest-neighbor (n.n.) repulsion.
Since such interactions are not contained in (1),
{\em 
charge  correlations appear here as pure correlation effect.} 
Results are shown for $U/t=16$, corresponding to 
$t\sim 0.3$ eV and $U\sim 4.5$ eV\cite{U}.
\begin{figure}
\epsfxsize=6.0cm
\centerline{\epsffile[30 20 470 680]{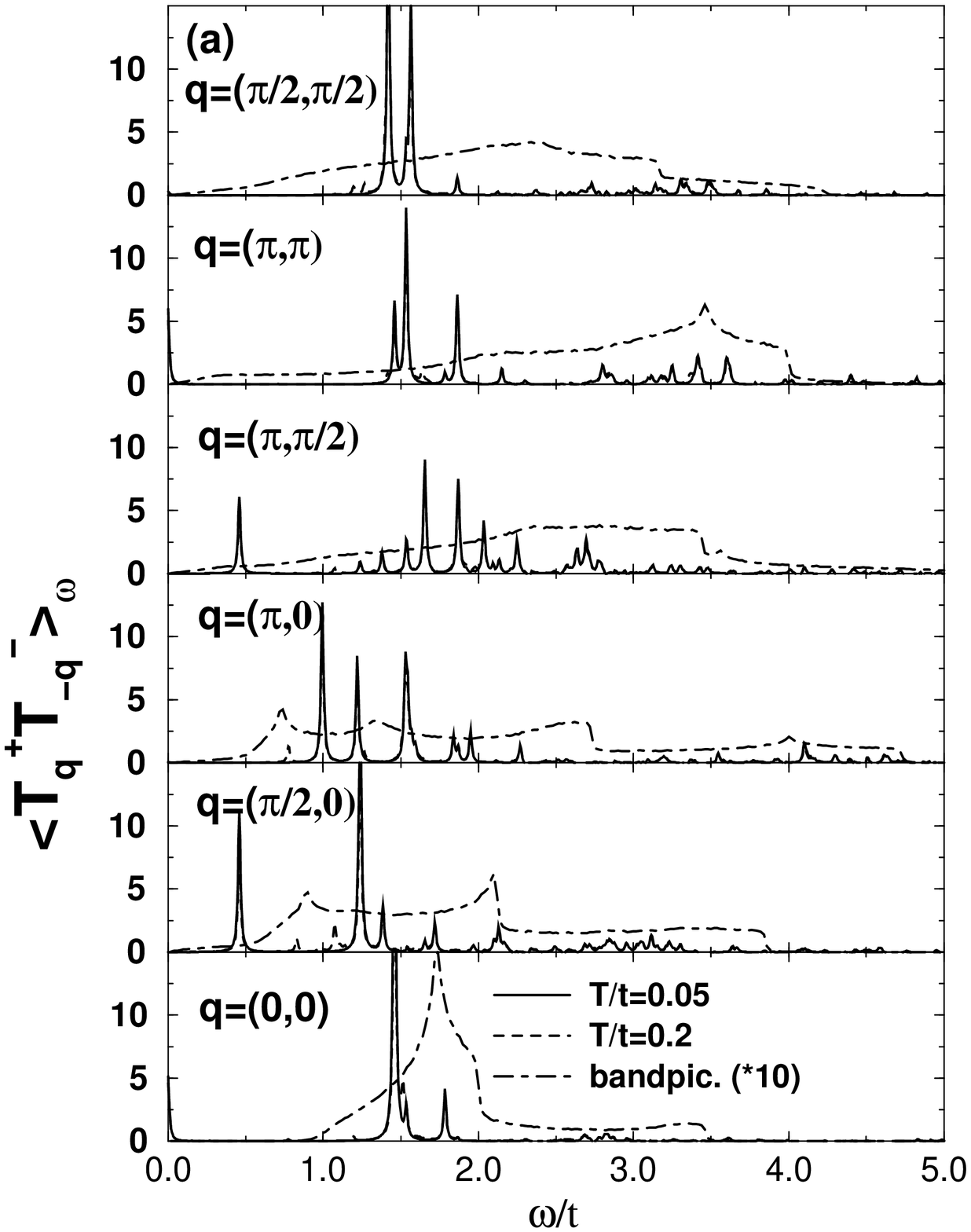}}
\epsfxsize=6.0cm
\centerline{\epsffile[30 20 470 680]{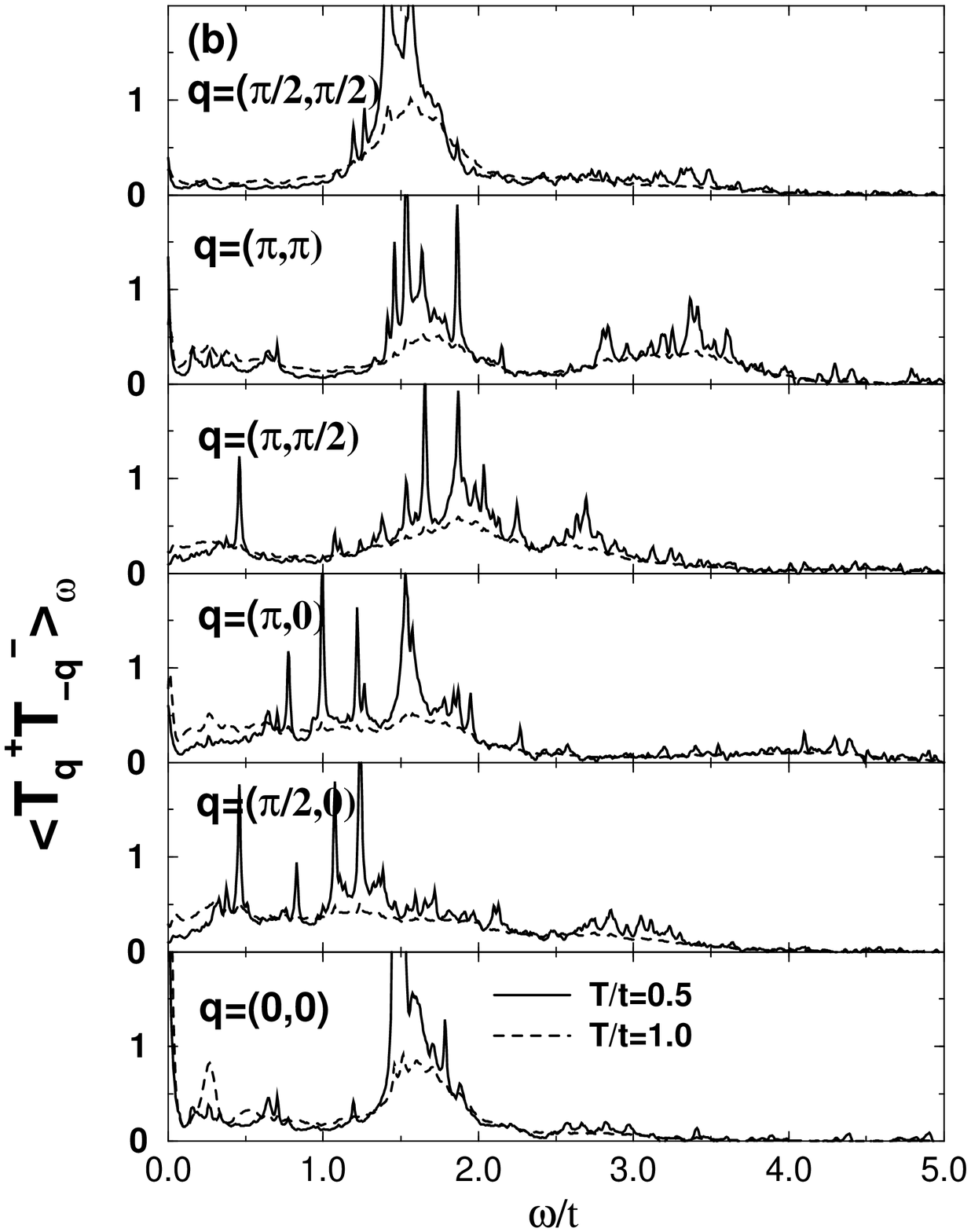}}
\noindent
\caption{ \label{fig2}
Transverse orbital excitation spectra 
$\langle T^+_{\bf q}T^-_{\bf -q}\rangle$  (a) in the orbital
ordered low-temperature regime
for various temperatures and (b) in the high-temperature regime 
with gapless orbital excitations for same parameters as in Fig. 1. 
Spectra are broadened using $\Gamma=0.01 t$.
In (a) a comparison with results from
the bandpicture is included (dash-dotted lines). }
\end{figure}
The orbital excitations are crucial
for the understanding of the optical conductivity of manganites 
in the FM phase. The form of the orbital excitation spectra 
changes significantly  with doping. 
In the undoped case with alternating orbital order due to the interaction (3)
the orbital excitation spectra 
$\langle T^+_{\bf q}T^-_{\bf -q}\rangle$
are essentially described by two collective modes\cite{Brink99},
while at low doping with the emergence of  $x^2$-$y^2$ orbital order
this changes into a single collective orbital mode
which is gapped.
In the high doping case, $x\sim 0.6$, studied here (Fig. 2a) the single mode features
have almost disappeared and the small cluster results show the features of a 
continuum with spectral weight distributed up to energies of about $\sim 4t$.
We note, however, that the dominant peaks in these spectra 
correspond to the collective excitations observed for systems at low doping.
\newline\indent
In the low and moderate doping regime a slave fermion approach provides a 
satisfactory description of the orbital excitations.
In such a treatment holes are described by spinless fermions $h_{\bf i}$
and the orbital pseudospins by Schwinger boson operators $b_{{\bf i} a}$
leading to the kinetic energy Hamiltonian 
$H_t=-\sum_{\langle {\bf i j}\rangle a b} t_{\bf ij}^{ab} h_{\bf i}h_{\bf j}^{\dagger}
{b}_{{\bf i} a}^{\dagger} {b}_{{\bf j} b}+ H.c.$
Subsequent mean-field treatment, i.e. assuming condensation of $b_{{\bf i} \uparrow}$,
plus Gaussian fluctuations leads to 
$H\sim \sum \omega_{\bf q} b_{\bf q}^{\dagger} b_{\bf q}$ with
$ \omega_{\bf q}= 3xt(1-\gamma_{\bf q}/3)$, where 
$\gamma_{\bf q}=(\cos q_x+\cos q_y)/2$.
This result tells us that the energy scale of orbital excitations is proportional
to the doping concentration $x$, which is consistent with the diagonalization
data. The scaling of $ \omega_{\bf q}$ with $x$ is due to the fact that 
$x^2$-$y^2$ {\em orbital order is induced by the kinetic energy of the holes}
which overcomes the superexchange interaction (3). 
Further boson-fermion coupling terms not included in this consideration lead 
to deviations from single mode behavior and will be discussed elsewhere.
\newline\indent
In this work we provide a comparison with an analytical approach based 
on the band picture which was proposed by Shiba and coworkers
\cite{Shiba97}.  
This Fermi-liquid approach assumes that  $U/t$ is not too large in
sytems like La$_{1-x}$Sr$_x$MnO$_3$.
Moreover the approach is expected to be useful in the 
dilute ($e_g$ electron density) limit. 
In this case the constraint in (1) may be
neglected and correlation effects can be included in a perturbative way.
Thus the kinetic energy can be diagonalized, leading to two tight-binding
(TB) bands. For the planar model the energies are 
$E_{\pm}/t = 2\gamma_{\bf k} \pm \sqrt{\gamma^2_{\bf k}+3\eta^2_{\bf k}}-\mu$,
where $\mu$ is the chemical potential and $\eta_{\bf k}=(\cos k_x-\cos k_y)/2$ .
Results for the transverse orbital excitation spectrum
$\langle T^+_{\bf q}T^-_{\bf -q}\rangle$ are shown in Fig. 2a for $T=0$ 
and compared with the diagonalization data.
The TB-spectra for the planar model, i.e. neglecting correlation effects, 
show a broad continuum with several 
van Hove singularities. The pronounced structures in the diagonalization 
data around $\omega\sim 3xt$, reminescent of the collective orbital
waves at lower doping, are not reproduced.  
\newline\indent
At temperatures $T>0.2$ thermal fluctuations destroy the orbital order (Fig. 1).
In this regime the orbital excitation gap changes into a pseudo gap with
increasing low energy spectral weight (Fig. 2b). The appearance of low
energy excitations in the orbital liquid phase has a dramatic effect on 
the optical conductivity as we shall see in the following.
\newline\indent
The frequency dependent conductivity consists of two parts
\cite{Shastry90},  
$\sigma_0(\omega)=2\pi e^2 D_c\delta(\omega) + \sigma(\omega)$.
The $\delta$-function contribution is proportional to the charge
stiffness $D_c$, which vanishes in insulators. 
The finite frequency absorption (or regular part) $\sigma(\omega)$
is determined by the current-current correlation function:
\begin{equation}
\sigma(\omega)=\frac{1-e^{-\omega/T}}{N\omega} Re \int_0^{\infty}dt
e^{i\omega t}\langle j_x(t)j_x\rangle.
\end{equation}
where the
$x$-component of the current operator is given by
$j^{(1)}_x=-ie \sum_{{\bf j+u} a b} t^{a b}_{\bf j+u, j}\; u_x \;
{\tilde d}^{\dagger}_{{\bf j+u} a} {\tilde d}_{{\bf j} b}$.
\begin{figure}
\epsfxsize=5.0cm
\centerline{\epsffile[-20 20 450 680]{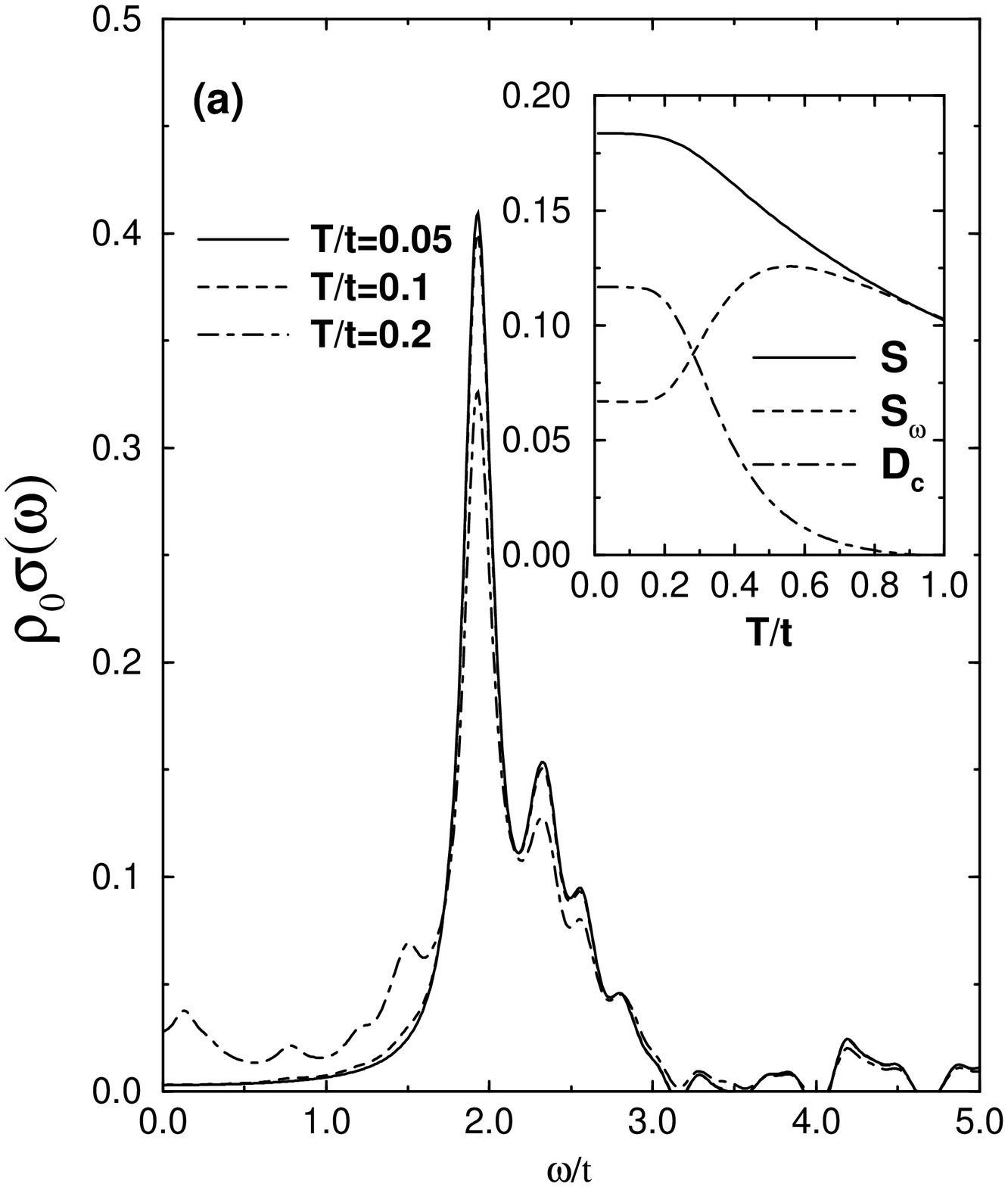}}
\epsfxsize=5.0cm
\centerline{\epsffile[-40 20 430 680]{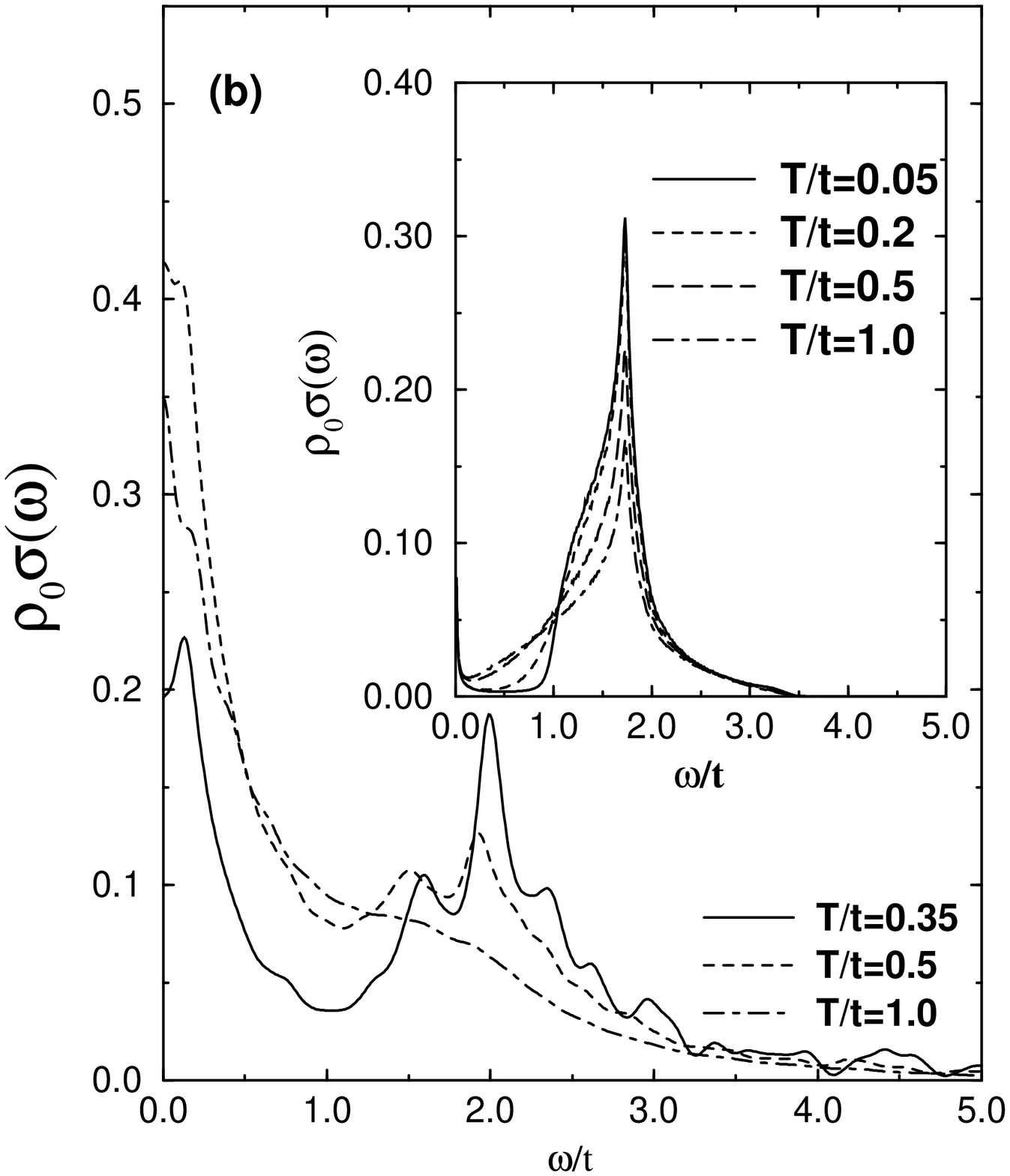}}
\noindent
\caption{ \label{fig3}
Optical conductivity (a) in the $x^2$-$y^2$ orbital ordered low-temperature phase
for $T=0.05, 0.1$ and $0.2 t$ and (b) in the high-temperature regime 
with gapless orbital excitations. Parameters as in Fig. 1
and $\Gamma=0.1 t$.   
The inset in (a) shows the Drude weight $D_c$,
the sum rule $S$ and 
the incoherent spectral weight $S_{\omega}$ versus temperature. 
Inset (b): Optical conductivity in the band model for the hole concentration $x=0.6$
for various temperatures ($\Gamma=0.02 t$).
}
\end{figure}
Figure 3a shows the regular part $\sigma(\omega)$ at low temperature
in the orbital ordered phase (the Drude peak $\sim D_c \delta(\omega)$
is not shown). 
The conductivity is given in dimensionless form $\rho_0\sigma(\omega)$, 
where $\rho_0=\hbar a /e^2$ \cite{Horsch98}. 
A typical value is $\rho_0\sim0.2\cdot 10^{-3}\; \Omega$cm.
The three data sets for $T=0.05 t$, $0.1 t$ and $0.2 t$ almost coincide,
consistent with the saturation of orbital order below $T=0.2 t$ (Fig. 1).
The finite frequency absorption $\sigma(\omega)$ 
for $x=0.6$ in Fig. 3a shows a well developed gap
$\omega_{opt}\sim 2 t$, which is due to the 
pronounced gap in the orbital excitation spectra. 
We note that in the $x^2$-$y^2$ ordered phase at low doping studied in 
Ref.\cite{Horsch98}
there is also a pseudogap in $\sigma(\omega)$, which is however smaller in 
size and only observable at low temperatures, because of the smaller
orbital excitation gap.
The inset in Fig. 3a shows the temperature dependence of the optical sum rule  
$S=-\frac{1}{2N}\langle H^{kin}_{xx}\rangle$, the absorption at
finite frequency
$S_{\omega}=\frac{1}{\pi e^2} \int^{\infty}_{0^+}\sigma(\omega)d\omega$
and the Drude weight $D_c$.
The latter quantity is determined from the sum rule $S=D_c+S_{\omega}$\cite{Horsch98}.
The weight of the Drude peak strongly increases below $T=0.5 t$ and 
saturates below $T=0.2 t$, i.e. similar to the evolution of orbital order
in Fig. 1.
At low temperature about 60$\%$ of the spectral weight is in the Drude peak, 
which may however depend somewhat on the cluster size.
\newline\indent
There is a dramatic shift of spectral weight in
$\sigma(\omega)$ at higher temperatures
when entering the orbital liquid phase (Fig. 3b).
The conductivity is characterized by a broad peak centered at $\omega=0$,
reminescent of a conventional Drude peak with a natural broadening $\Gamma\sim 0.5 t$.
We attribute this intrinsic broadening to the finite density of low-energy
orbital excitations in the orbital liquid phase (Fig. 2b). 
There is still some structure in $\sigma(\omega)$ near $\omega\sim 2 t$
close to the onset of absorption in the ordered low-$T$ phase.
\newline\indent
The conductivity for the planar model calculated using the band picture
is shown in Fig. 3b (inset). 
In this approach the finite frequency absorption is due to interband
scattering, i.e. a consequence of the off-diagonal hopping $t^{\uparrow\downarrow}$.
At low temperature the results for $\sigma(\omega)$
obtained by neglecting correlation effects are not very different from the
exact diagonalization result. Both methods yield a clear gap and the main 
absorption is close to $\omega\sim 2 t$. The onset of absorption in the band
picture is near $\omega\sim 1 t$, i.e. about $0.5 t$ lower than in the 
cluster calculation.
The temperature dependence in the band picture is, however, quite distinct.
The filling in of spectral weight at low energy occurs only gradually, i.e.
even at temperatures as high as $T=1 t$ there is only little low-energy
spectral weight.
The temperature scale of the transition from the $x^2$-$y^2$ ordered phase 
to the orbital liquid is a consequence of strong correlations, and not contained 
in the band picture, where the natural energy scale is much larger, i.e. 
given by the band width $W=6 t$.  
\newline\indent
Planar $x^2$-$y^2$ correlations are favored by the kinetic energy of holes,
since the partially filled $x^2$-$y^2$ band has the largest overlap.
The motion is quasi-2D since the 
hopping along the c-direction vanishes by symmetry.
Moreover the antiferromagnetic coupling between layers in the A-phase in combination 
with double exchange would prevent coherent c-axis motion.
\newline\indent
In a cubic system there are now two possibilities:
(i) Since $x^2$-$y^2$ correlations are not preferred compared to  $y^2$-$z^2$
and $z^2$-$x^2$ correlations in the corresponding perpendicular planes, 
the system remains in a disordered quantum liquid state. This is the
orbital liquid state proposed by Nagaosa and coworkers\cite{Ishihara97b}.
(ii) Another possibility is spontaneous symmetry breaking selecting one 
particular set of planes, e.g. the formation of $x^2$-$y^2$ long-range order.
We note that in the planar model the cubic symmetry is explicitely broken,
therefore case (i) does not occur for this geometry. In the planar model
the orbital liquid develops at higher temperature due to
thermal fluctuations.
The orbital ordering temperature  $T_c^{oo}$ increases with doping.
\newline\indent
We believe that in cubic manganites both cases are realized in different
doping regimes. At moderate doping concentration the orbital liquid state 
appears to be the proper ground state in combination with the isotropic
ferromagnetic spin structure. This is consistent with the broad absorption
in $\sigma(\omega)$ observed by Okimoto {\it et al.} and others. These 
spectra can be compared with our result for the orbital liquid in the
planar model (Fig. 3b) which is stabilized by thermal fluctuations.
\newline\indent
The observation of the A-phase at high doping concentration by 
Kawano {\it et al.}\cite{Kawano97} and Akimoto {\it et al.}\cite{Akimoto98} 
seems to be a realization of 
case (ii) where symmetry is broken spontaneously. Our prediction
for this phase are gapped orbital excitations with energy $\sim 3xt$.
The optical conductivity of this quasi-2D metallic state is expected to consist 
of a large Drude peak, while the finite frequency absorption shows a pronounced
gap due to the gapped orbital excitations.  

%

\end{multicols}

\begin{references}
\bibitem{Kawano97}
H. Kawano {\it et al.}, Phys. Rev. Lett. {\bf 78}, 4253 (1997).

\bibitem{Akimoto98}
T. Akimoto {\it et al.}, Phys. Rev. B {\bf 57}, R5594 (1998).

\bibitem{Murakami98}
Y. Murakami {\it et al.}, Phys. Rev. Lett. {\bf 81}, 582 (1998).

\bibitem{Ishihara97b}
S. Ishihara, M. Yamanaka, and N. Nagaosa, Phys. Rev. B {\bf 56}, 686 (1997).

\bibitem{Kilian98}
R. Kilian and G. Khaliullin, Phys. Rev. B {\bf 58}, R11841 (1998).


\bibitem{Okimoto95}
Y. Okimoto {\it et al.}, Phys. Rev. Lett {\bf 75}, 109 (1995) and
Phys. Rev. B {\bf 55}, 4206 (1997).

\bibitem{Kim98}
K. H. Kim, J. H. Jung, and T. W. Noh,  Phys. Rev. Lett {\bf 81}, 1517 (1998).

\bibitem{Quijada98}
M. Quijada {\it et al.}, Phys. Rev. B {\bf 58}, 16093 (1998).

\bibitem{Shiba97}
H. Shiba, R. Shiina, and A. Takahashi, J. Phys. Soc. Jpn. {\bf 66},
941 (1997).
A. Takahashi and H. Shiba, Eur. Phys. J. B {\bf 5}, 413 (1998).

\bibitem{Horsch98}
P. Horsch, J. Jakli\v c, and F. Mack, Phys. Rev. B {\bf 59}, 6217 (1999).

\bibitem{Ishihara97a}
S. Ishihara, J. Inoue, and S. Maekawa, Phys. Rev. B {\bf 55}, 8280 (1997).

\bibitem{U}
The effective $U$ is $U$-$J_H$ for an $e_g$ model, and $U$-$5 J_H$ for a full
d-band model, see A. M. Ole\'s and L. F. Feiner, J. Superconductivity 
{\bf 12}, 299 (1999).

\bibitem{Jaklic94}
J. Jakli\v c and P. Prelov\v sek, Phys. Rev. B {\bf 49}, 5065 (1994);
ibid {\bf 50}, 7129 (1994); ibid {\bf 52}, 6903 (1995).

\bibitem{BC}
Twisted boundary conditions are used as in Ref.\cite{Horsch98}.

\bibitem{Brink99}
J. van den Brink, P. Horsch, F. Mack, and A. M. Ole\'s, 
Phys. Rev. B {\bf 59}, 6795 (1999).

\bibitem{Shastry90}
B. S. Shastry and B. Sutherland, Phys. Rev. Lett. {\bf 65}, 243 (1990).

\end{references}
\end{document}